\documentclass[a4paper]{jpconf}
\usepackage{graphicx}
\begin{document}
\title{Mott insulating state in a quarter-filled two-orbital Hubbard chain 
with different bandwidths}

\author{Satoshi Miyashita$^{1,2}$, Yasufumi Yamashita$^{2,3}$, 
Kenji Yonemitsu$^{2,3}$, Akihisa Koga$^{4}$, and Norio Kawakami$^{4}$}

\address{$^1$Institute for Materials Research, Tohoku University, 
Sendai 980-8577, Japan}
\address{$^2$Institute for Molecular Science, 
Okazaki 444-8585, Japan}
\address{$^3$Department of Functional Molecular Science, Graduate University 
for Advanced Studies, Okazaki 444-8585, Japan}
\address{$^4$Department of Physics, Kyoto University, 
Kyoto 606-8502, Japan}

\ead{satoshi@ims.ac.jp}

\begin{abstract}
We investigate the ground-state properties of the one-dimensional two-band Hubbard model with different bandwidths. The density-matrix renormalization group method is applied to calculate the averaged electron occupancies $n$ as a function of the chemical potential $\mu$. Both at quarter and half fillings, ``charge plateaux'' appear in the $n$-$\mu$ plot, where $d\mu/dn$ diverges and the Mott insulating states are realized. 
To see how the orbital polarization in the one-quarter charge plateau develops, we apply the second-order perturbation theory from the strong-coupling limit at quarter filling. The resultant Kugel-Khomskii spin-orbital model includes a $magnetic$ field coupled to orbital pseudo-spins. This field originates from the discrepancy between the two bandwidths and leads to a finite orbital pseudo-spin magnetization.
\end{abstract}

%%%%%%%%%%%%%%%%%%%%%%%%
%%%%%%%%%%%%%%%%%%%%%%%%
%%%%  Introduction  %%%%
%%%%%%%%%%%%%%%%%%%%%%%%
%%%%%%%%%%%%%%%%%%%%%%%%
%\section{Introduction}
``Orbital degrees of freedom'' has been one of the important keywords to understand the low temperature properties of strongly correlated electron systems. Among them, recently, the orbital-selective Mott transition (OSMT) is proposed to explain the exotic heavy metallic state in Ca$_{2-x}$Sr$_x$RuO$_4$\cite{Nakatsuji,Anisimov}. 
To illustrate this scenario, the two-orbital model with different bandwidths has been extensively investigated by dynamical mean-field theory\cite{Liebsch, Koga}. They have clarified the realization conditions of the OSMT by changing various parameters, such as Coulomb interaction parameters, band structures, band filling, and others. 
In this paper, we also study the ground state properties of the two-band Hubbard model with different bandwidths, but in one dimension (1D). Our main concern here is whether the OSMT survives in 1D, where the effect of quantum fluctuations is most severely enhanced.

%
%%%%%%%%%%%%%%%%%
%%%%%%%%%%%%%%%%%
%%%%  Model  %%%%
%%%%%%%%%%%%%%%%%
%%%%%%%%%%%%%%%%%
%\section{Model and Method}
First, let us begin with the description of the 1D two-orbital Hubbard model\cite{Sakamoto} defined by;
%%%%%%%%%%%%%%%%%%
\begin{eqnarray}
{\cal H} &=& \sum_{ij\alpha\sigma}  \Big\{-t_{\alpha}(\delta_{i+1,j}+\delta_{i-1,j}) - \mu \delta_{ij} \Big\}
              c^\dag_{i\alpha\sigma} c_{j\alpha\sigma}
          +U \sum_{i\alpha} n_{i\alpha\uparrow} n_{i\alpha\downarrow}
          +U'\sum_{i\sigma\sigma'} n_{i1\sigma} n_{i2\sigma'}\nonumber \\
         &&
          -J \sum_{i\sigma\sigma'}c^\dag_{i1\sigma} c_{i1\sigma'}c^\dag_{i2\sigma'} c_{i2\sigma}
          -J\sum_{i}\left( c^\dag_{i1\uparrow} c^\dag_{i1\downarrow}c_{i2\uparrow} c_{i2\downarrow} +h.c. \right)\;, 
\label{dege-Hub}
\end{eqnarray}
%%%%%%%%%%%%%%%%%%
where ${\it c}_{j \alpha \sigma}^{\dag}$ creates an electron with spin $\sigma$ (=$\uparrow$ or $\downarrow$) and orbital ${\it \alpha}$ (= 1 or 2) at the $j$-th site. The electron transfers of strength $t_{\alpha}$ are possible between the same type of neighboring orbitals and $\mu$ denotes the chemical potential. As for the intra-site Coulomb interactions, we assume the rotational invariance for simplicity, and thus the Coulomb parameters, $U$, $U'$, and $J$, always hold the following relation; $U=U'+2J$. As long as the intra-atomic Coulomb interactions are concerned, physically relevant parameters would satisfy $U>U'>J$. However, we do not restrict our calculations to this parameter region, since in general the above Hamiltonian can be also viewed as a variation of coupled Hubbard chains.

In this paper, we have used the infinite-size density-matrix renormalization group (DMRG) algorithm\cite{White1,White2} modified by adapting the recursion relation to the wave function, namely, the wave function is so upgraded that it should gradually approach the ground-state wave function when the system size is extended. This technique, called the product-wave-function renormalization-group method\cite{Nishino,Hieida}, enhances the accuracy of calculations and overcomes metastability problems during the renormalizing process if the ground state to target is gapless. 

%%%%%%%%%%%%%%%%%%%%%%%%%%%%%
%%%%%%%%%%%%%%%%%%%%%%%%%%%%%
%%%%  Numerical results  %%%%
%%%%%%%%%%%%%%%%%%%%%%%%%%%%%
%%%%%%%%%%%%%%%%%%%%%%%%%%%%%
%
%
%%%%%%%%%%%%%%%%%%%%%%%%%%%%%%%%%%%%%%%%%%%%%%%%%%%%%%%%%%%%%%%%%%%%%%
%%% Fig.1            %%%
%%% U=10,U'=5, J=2.5 %%%
%%%%%%%%%%%%%%%%%%%%%%%%
\begin{figure}[htb]\begin{center}\includegraphics[width=70mm]{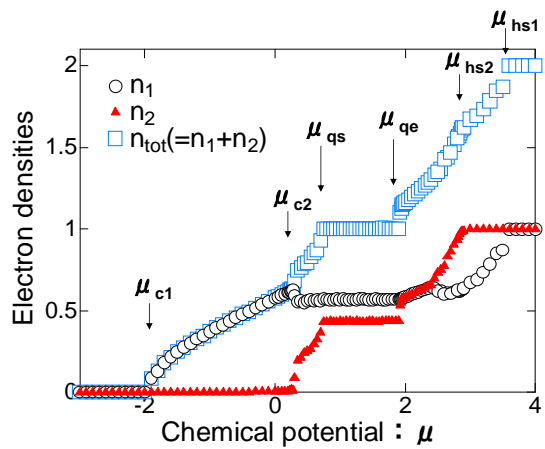}\end{center}\caption{(Color online) 
The averaged electron densities of orbital 1 ($n_1$, denoted by open circles), orbital 2 ($n_2$, solid triangles) and their sum ($n_{\rm tot}$, open squares) as a function of chemical potential $\mu$. Coulomb interactions are set to be $U=10$, $U'=5$ and $J=2.5$.}\label{f1}
\end{figure}
%%%%%%%%%%%%%%%%%%%%%%%%%%%%%%%%%%%%%%%%%%%%%%%%%%%%%%%%%%%%%%%%%%%%%%
%
%%%%%%%%%%%%%%%%%%%%%%%%%%%%%%%%%%%%%%%%%%%%%%%%%%%%%%%%%%%%%%%%%%%%%%
%%% Fig.2                       %%%
%%% U=10,J=(U-U')/2, 1/4-filled %%%
%%%%%%%%%%%%%%%%%%%%%%%%%%%%%%%%%%%
\begin{figure}[htb]\begin{center}\includegraphics[width=70mm]{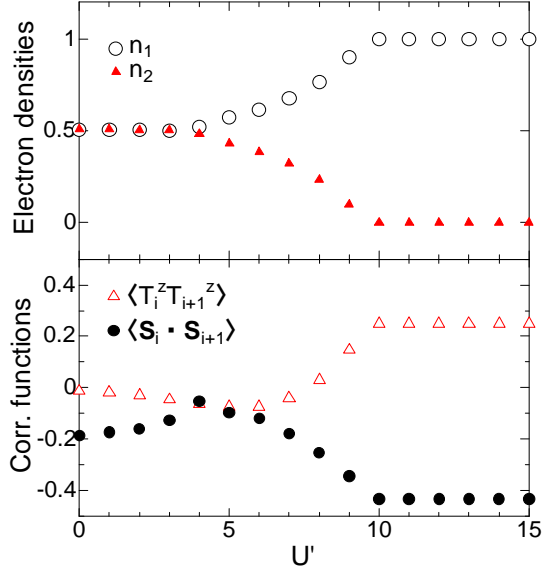}\end{center}\caption{(Color online) 
The averaged electron densities, $n_1$ and $n_2$ (upper panel), and the nearest neighbor spin-spin and orbital-orbital correlation functions (lower panel) as a function of $U'$, where $n_{\rm tot}=1$, $U=10$ and $U=U'+2J$ holds.}\label{f2}
\end{figure}
%%%%%%%%%%%%%%%%%%%%%%%%%%%%%%%%%%%%%%%%%%%%%%%%%%%%%%%%%%%%%%%%%%%%%%
%
%\section{Numerical results and discussions}
The different bandwidths are implemented in Eq.(\ref{dege-Hub}) by taking $t_1=1$ and $t_2=0.5$. Since we are mainly interested in strongly correlated regime, $U$ is always fixed at 10 throughout this paper and $U'$ as well as $J$ are varied while keeping $U=U'+2J$. By means of the DMRG technique, we have calculated the electron density of $\alpha$-orbital defined by $n_{\alpha}=\sum_{i\sigma}\langle n_{i\alpha\sigma}\rangle/N$, where $N$ is the total number of sites. Figure 1 shows the $\mu$-dependence of $n_1$, $n_2$, and $n_{\rm tot}\equiv n_1+n_2$ for $U'=5$ and $J=2.5$. 
Let us take a closer look at Fig. \ref{f1} from the low chemical potential region.
When $\mu$ exceeds $\mu_{c1}=-2t_1$, the bottom of the orbital-1 energy band, $n_1$ increases sharply while $n_2$ stays zero until $\mu$ reaches $\mu_{c2}$. At $\mu_{c2}$, $n_1$ and $n_2$, respectively, drops down and jumps up suddenly. Then between $\mu_{c2}$ and $\mu_{qs}$, $n_1$ ($n_2$) shows a gradual (steep) monotonic increase and, finally, the first plateaux for both $n_1$ and $n_2$ appear simultaneously between $\mu_{qs}$ and $\mu_{qe}$. At these plateaux, the values of $n_1$ and $n_2$ are irrational but $n_{\rm tot}$ definitely equals to unity corresponding to one-quarter filling. Therefore, this Mott insulating state itself is not controversial except for that $n_1$ is not equal to $n_2$. We will come back to this point later.
With further increase of $\mu$ above $\mu_{qe}$ in Fig. \ref{f1}, we find the next plateau at $n_2=1$ between $\mu_{hs2}$ and $\mu_{hs1}$, which means the charge gap opens for the orbital-2 band. On the other hand, $n_1$ increases continuously toward unity in this region and thus the orbital-1 band should be metallic. This implies that the filling-control OSMT takes place at $\mu_{hs2}$. Then, the orbital-selective Mott state with one orbital localized and the other itinerant is realized just below one-half filling in the present system. For $\mu\ge \mu_{hs1}$, the $n_1=1$ plateau emerges in addition to the $n_2=1$ plateau, which correspond to the normal half-filled Mott insulating state.

%%%%%%%%%%%%%%%%%%%%%%
%
Hereafter, we will investigate the quarter-filled Mott state in more detail. For this purpose, the DMRG calculations are performed within the $S^z_{\rm tot}\equiv\sum_iS_i^z=0$ subspace at quarter filling. In Fig. \ref{f2} shown are the electron densities ($n_1$ and $n_2$, upper panel) and correlation functions ($\langle T_i^z T_{i+1}^z \rangle$ and $\langle {\vec S}_i \cdot {\vec S}_{i+1} \rangle$, lower panel) as a function of $U'$ for $U=10$ and $J=(U-U')/2$. 
Considering the atomic limit, there are three possible ground states characterized by;
i) spin-triplet and orbital-singlet states like $c^{\dagger}_{i1\uparrow}c^{\dagger}_{i2\uparrow}|0\rangle$ (averaged site energy $\varepsilon=(U'-J)/2=(3U'-U)/4$), ii) singly occupied states like $c^{\dagger}_{i\alpha\sigma}|0\rangle$ (null site energy), and iii) spin-singlet and orbital-triplet states like $(c^{\dagger}_{i1\uparrow}c^{\dagger}_{i1\downarrow}+c^{\dagger}_{i2\uparrow}c^{\dagger}_{i2\downarrow})|0\rangle$ ($\varepsilon=(U+J)/2=(3U-U')/4$). Therefore, two quantum critical points, $U'_{c1}=U/3$ and $U'_{c2}=3U$, separate the above three phases when $t_1=t_2=0$. The singularity around $U'=3.5$ in Fig. \ref{f2} seems to correspond to $U'_{c1}$ and, in fact, the vanishing orbital polarization below $U'\le 3.5$ is consistent with the formation of a local orbital singlet. With further increase of $U'$ above $U'_{c1}$, an orbital magnetization: $\langle T_i^z \rangle \equiv (n_1-n_2)/2$ develops accompanied by the antiferromagnetic spin-spin and the ferromagnetic orbital-orbital correlations. These results show that, at the one-quarter plateau, the magnitude of charge disproportionation between $n_1$ and $n_2$ depends on Coulomb interaction parameters and only the total electron density is preserved. For $U'\ge U=10$, $\langle T_i^z \rangle$ is saturated and all electrons reside in the orbital 1. In such a strong-coupling regime, the system should be identical to the isotropic Heisenberg spin chain, which is confirmed by $\langle{\vec S}_i\cdot{\vec S}_{i+1}\rangle=1/4-\ln{2}\sim-0.443$ as well as $\langle T_i^z T_{i+1}^z \rangle=1/4$ in Fig. \ref{f2}.

%%%%%%%%%%%%%%%%%%%%%%%
%%%%%%%%%%%%%%%%%%%%%%%
%%%%  Discussions  %%%%
%%%%%%%%%%%%%%%%%%%%%%%
%%%%%%%%%%%%%%%%%%%%%%%
%\section{Discussions}
In order to examine how the orbital polarization at quarter filling is developed with increasing $U'$, we have derived the effective spin-orbital Hamiltonian of Kugel-Khomskii type\cite{Kugel}. Starting from the atomic limit and within $U'_{c1}\le U'\le U'_{c2}$, the effective Hamiltonian is given by; ${\cal H}_{\rm eff}=\sum_i\left[{\cal H}_{i,i+1}^{(so)} - \Delta_{i,i+1}^{(S)} (T_i^z + T_{i+1}^z)\right]$ with;
%%%%%%%%%%%%%%%%%%
\begin{eqnarray}
%{\cal H}_{\rm eff} &=& \sum_{<i,j>}\left[{\cal H}_{ij}^{so} - h_{ij}^T (T_i^z + T_j^z)\right]\;,\label{eff-Hub}
{\cal H}_{i,j}^{(so)} &=&
%          &=& \sum_{<i,j>} 
%              \left[
               -\left( 
                 \frac{1}{U-J} + \frac{1}{U+J} 
                \right)
                \left(
                 \frac{1}{4} - {\vec S}_i \cdot {\vec S}_{j}
                \right)
%               \right.
%\nonumber \\ 
%          && \ \ \ \ \ \ \ \ \ \ \ 
%          \times
                \left\{
                  2( {t_1}^2 + {t_2}^2 ) 
                   \left( \frac{1}{4}+T_i^zT_{j}^z \right)
%                  2{t_1}^2 \left( \frac{1}{2}+T_i^z \right)
%                           \left( \frac{1}{2}+T_{j}^z \right)
%                 +2{t_2}^2 \left( \frac{1}{2}-T_i^z \right)
%                           \left( \frac{1}{2}-T_{j}^z \right)
                \right\}
\nonumber \\
          && 
%\ \ \ \ \ \ \ \ 
               -\left( 
                 \frac{1}{U-J} - \frac{1}{U+J} 
                \right)
                \left(
                 \frac{1}{4} - {\vec S}_i \cdot {\vec S}_{j}
                \right)
%\nonumber \\ 
%          && \ \ \ \ \ \ \ \ \ \ \ 
%          \times
                \left\{
                 2t_1 t_2 \left( T_i^+T_{j}^+ + T_i^-T_{j}^- \right)
                \right\}
\nonumber \\
          && 
%\ \ \ \ \ \ \ \ 
               -\frac{1}{U'+J}
                \left(
                 \frac{1}{4} - {\vec S}_i \cdot {\vec S}_{j}
                \right)
%\nonumber \\ 
%          && 
%\ \ \ \ \ \ \ \ \ \ \ 
%          \times
                \left\{
                  2\left( {t_1}^2+{t_2}^2 \right)
                   \left( \frac{1}{4}-T_i^zT_{j}^z \right)
                 +2t_1 t_2
                   \left( T_i^+T_{j}^- + T_i^-T_{j}^+ \right)
                \right\}
\nonumber \\
          && 
%\ \ \ \ \ \ \ \ 
               -\frac{1}{U'-J}
                \left(
                 \frac{3}{4} + {\vec S}_i \cdot {\vec S}_{j}
                \right)
%\nonumber \\ 
%          && 
%\ \ \ \ \ \ \ \ \ \ \ 
%          \times
%               \left.
                \left\{
                  2\left( {t_1}^2+{t_2}^2 \right)
                   \left( \frac{1}{4}-T_i^zT_{j}^z \right)
                 -2t_1 t_2
                   \left( T_i^+T_{j}^- + T_i^-T_{j}^+ \right)
                \right\}
%               \right]
\;,
\nonumber\\
\Delta_{i,j}^{(S)} &=&  \left(
               {t_1}^2 - {t_2}^2
              \right)
              \left(
               \frac{1}{U-J} + \frac{1}{U+J}
              \right)
              \left(
               \frac{1}{4} - {\vec S}_i \cdot {\vec S}_{j}
              \right)
\nonumber
\;,
\end{eqnarray}
%%%%%%%%%%%%%%%%%%
where the spin-1/2 ($\vec S$) and orbital pseudospin-1/2 ($\vec T$) operators are, respectively, defined by 
${\vec S}_j{\equiv}\sum_{\alpha\sigma\sigma'}\left(c_{j\alpha\sigma}^\dag {\vec \tau}_{\sigma \sigma'} c_{j\alpha\sigma'}\right)/2$ and 
${\vec T}_j{\equiv}\sum_{\sigma\alpha\alpha'}\left(c_{j\alpha\sigma}^\dag {\vec \tau}_{\alpha \alpha'} c_{j\alpha'\sigma}\right)/2$
with the use of the Pauli matrices $\vec \tau$'s.
$\Delta_{i,j}^{(S)}$ is an effective crystal field coupled to orbital pseudospins, which originates from the discrepancy between the two bandwidths, that is, $t_1\ne t_2$. 
%Since $\Delta^{(S)}$ is proportional to ${(3U-U')^{-1}}$, positive $\langle T_i^zT_{i+1}^z\rangle$ is induced with an increase of $U'$ ($U'<U'_{c2}$). 
Since the ${(U+J)^{-1}}$ term in $\Delta_{i,j}^{(S)}$, which is proportional to ${(3U-U')^{-1}}$, increases with an increase of $U'$ ($U'<U'_{c2}$), a ferromagnetic $\langle T_i^zT_{i+1}^z\rangle$ is expected to be developed accordingly. 
Therefore, the different bandwidth in the present model is indispensable for producing the orbital polarization. On the other hand, when $t_1=t_2$ and $U=U'$, $\Delta_{i,j}^{(S)}$ vanishes and ${\cal H_{\rm eff}}$ is equivalent to the SU(4) spin-orbital model\cite{Sutherland,Yamashita}. In this model, the ground state is an SU(4) singlet ground state with $\langle T_i^z\rangle=0$ corresponding to $n_1=n_2$.

%%%%%%%%%%%%%%%%%%%%%%
%%%%%%%%%%%%%%%%%%%%%%
%%%%  Conclusion  %%%%
%%%%%%%%%%%%%%%%%%%%%%
%%%%%%%%%%%%%%%%%%%%%%
%\section{Conclusion}
To conclude, we have investigated the ground-state properties of the 1D two-orbital Hubbard model with different transfer integrals by the DMRG method. Two plateaux appear in the total electron density as a function of the chemical potential, which suggests the existence of Mott insulating states. The first insulating state ranging in $\mu_{qs}\le \mu\le\mu_{qe}$ corresponds to the quarter-filled Mott insulator and the charge disproportionation between two orbitals exists due to the different bandwidth of the two bands. 
On the other hand, in the second Mott insulating state for $\mu_{hs1}\le\mu$, the two bands are both half filled and thus the total electron density is at half filling. It is also found that the OSMT takes place just below half filling, where the charge gap exists in the narrower band and the wider band remains metallic. The detailed analysis around the charge one-half plateau will be published elsewhere.

%%%%%%%%%%%%%%%%%%%%%%%%%%%
%%%%%%%%%%%%%%%%%%%%%%%%%%%
%%%%  Acknowledgments  %%%%
%%%%%%%%%%%%%%%%%%%%%%%%%%%
%%%%%%%%%%%%%%%%%%%%%%%%%%%
\section*{Acknowledgments}
This work was supported by the Next-Generation Supercomputer Project (Integrated Nanoscience) and Grant-in-Aid from the Ministry of Education, Culture, Sports, Science and Technology, Japan. A part of computations was done at the Supercomputer Center at the Institute for Solid State Physics, University of Tokyo.
% [17740226 (A. K.) and 19014013 (N. K.)]
%%%%%%%%%%%%%%%%%%%%%%
%%%%%%%%%%%%%%%%%%%%%%
%%%%  References  %%%%
%%%%%%%%%%%%%%%%%%%%%%
%%%%%%%%%%%%%%%%%%%%%%

\end{document}